# The Physics of Drying of Colloidal Dispersion – Pattern Formation and Desiccation Cracks

**Hisay Lama[1][†*], Ranajit Mondal[2][†]**

[1]*IBS Center for Soft and Living Matter, UNIST, Ulsan-44919, South Korea*

[2]*Polymer Engineering and Colloid Science Laboratory, Department of Chemical Engineering, IIT Madras, Chennai-600036, India*

**Abstract:** Drying of colloidal dispersion and their consolidation into a particulate deposit is a common phenomenon. This process involves various physical processes such as diffusion of liquid molecules into the ambient atmosphere and advection of dispersed particles via evaporation driven flow. The colloidal particles forming a dried deposit exhibits distinct patterns and frequently possess structural defects such as desiccation cracks. This chapter gives an introductory review of the drying of colloidal dispersion and various associated phenomena. In principle, the drying of colloid dispersion, the process of their consolidation and fluid-flow dynamics are all studied in numerous drying configuration. Here we explain drying induced phenomena concerning to the sessile drop drying. We begin with an introduction to colloids, provide background on the physics of drying and then explain the formation of pattern and the desiccation cracks. The role of evaporation driven flows and their influence on particle accumulation, the impact of various physical parameters on pattern formation and cracks are all briefly illustrated.

## INTRODUCTION

Colloid is a model system that has been widely explored to study various physical phenomena such as self-assembly, pattern formation and desiccation cracks [1]. It's a mixture that comprises of two phases - dispersed phase and continuous medium. Such a mixture with the particles as the dispersed phase and liquid as the continuous medium is commonly called colloidal dispersion. The size of dispersed particles in the continuous medium ranges from 10 nm to 10 μm. The dispersed particles in the medium experiences several body forces and interaction forces, such as hydrodynamic, gravitational, Brownian, van der Waals, electrostatic and capillary forces [1,2]. These body forces are exerted on the individual particles while the interaction forces are between the multiple particles. The interaction forces acting between these particles are either attractive or repulsive, for example, van der Waals and capillary forces are attractive while the electrostatic forces are repulsive. The relative strength of these interaction forces (van der Waals and electrostatic) determines the stability of a dispersion [2]. Note that the term colloid or colloidal dispersion are interchangeably used here, both refer to the particle dispersion.

[†‡]Contributed equally
*Corresponding author:hisaylama@gmail.com

The study of colloid has a lot of practical applications in various technological domains [1]. In this chapter, we focus only on the physics of drying, formation of dried pattern and the desiccation cracks.

Drying of colloid is a ubiquitous phenomenon that we see in daily life [1,3–5]. A common example is a coffee stain mark that forms after complete evaporation of fluid from a drop of the coffee [6,7]. Apart from drying of coffee, few other examples are - drying ink [8], drying of paint [9,10] and drying mud [11,12] etc. In general, the drying of complex fluids viz. particle dispersion, the polymer solution or biological fluids itself is a fascinating subject and has technological importance [5]. In principle, the process of drying involves an accumulation of particles, their rearrangement and evaporation of containing fluid. The process of development of colloid into a dried particle deposit is commonly believed to be dictated by the advection and diffusion processes that happen during drying. The particle deposit formed via evaporation driven processes is known to manifest into a plethora of patterns, accompanying structural defects such as desiccation cracks [3]. The particles forming the dried deposit are closely packed, sometimes also possessing self-assembled microstructure that results in an ordered crystal [13,14]. This ordered structure has potential application in the fabrication of optical devices.

Apart from all, the nucleation of defects such as desiccation cracks and their self-organization into various morphology has also become an active area of research [4]. The subject of desiccation cracks has received good attention for the past few decades and has become a popular subject of research among the diverse community of researcher. There are several research domains wherein the desiccation cracks are very popular, they are – geosciences [15], crackle-lithography [16–18], coating technology [3], paint restoration [10] and diagnostic [19]. Note that the formation of cracks in the deposit is undesirable for most of the practical applications but the possibility of tailoring them into a regular pattern has found potential applications in lithography [16,17]. This has further motivated the researcher's to manoeuvre multiple techniques for tailoring the cracks into the desired pattern.

†‡Contributed equally
*Corresponding author:hisaylama@gmail.com

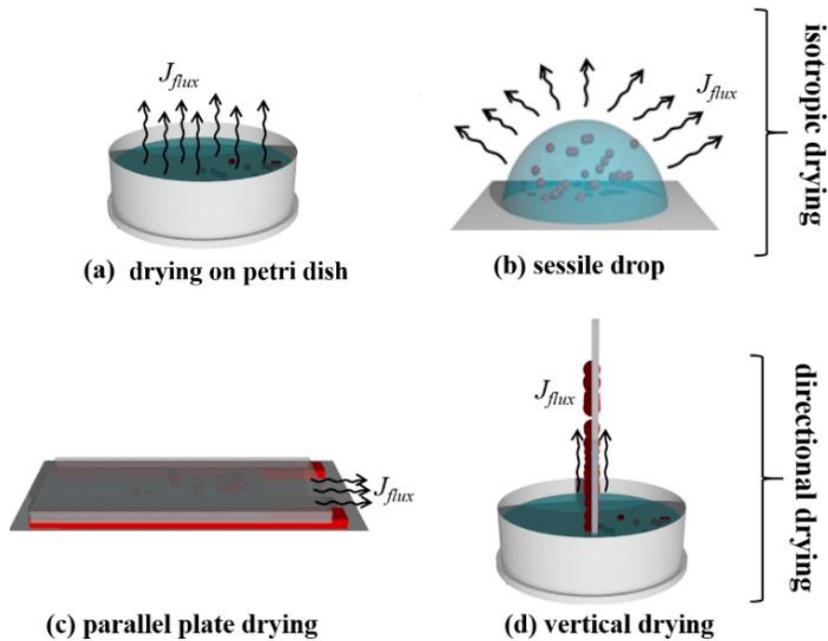

Figure 1: Schematic of various drying configuration namely, isotropic and directional, $J_{flux}$ points the evaporative flux. The drying of colloidal dispersion (a) on a Petri-dish, (b) in sessile-configuration, (c) inside the parallel plate and (d) via vertical drying configuration. The arrows indicate the direction of diffusion of vapour to an ambient atmosphere. (Images borrowed from PhD. thesis of Hisay Lama submitted to IIT Madras).

In the laboratory, the physics of drying and the pattern formation is commonly studied by conducting the model evaporation experiments [20], namely, isotropic or directional drying as shown in Fig. 1 (a)-(d). Drying colloids on Petri dish or microscope glass-slides can be classified as isotropic drying while drying inside a parallel plate or via vertical deposition can be classified as directional drying. Conducting experiments in these configurations allows one to theoretically model the dynamics of fluid flow and quantify several parameters, such as diffusion coefficient and evaporative flux associated with the drying process. Among all of these aforementioned drying configurations, sessile drop configuration has got more attention due to its ubiquity. The sessile configuration has an axisymmetric geometry and quantities that characterize such drying phenomena can easily be extracted in terms of 3D-cylindrical polar coordinates. Drying in sessile configuration has been widely exploited to understand the physics of drying and investigate the drying related phenomenon. Therefore, we will also describe the drying related phenomenon mostly for sessile drop configuration.

In this chapter, first, we discuss the phenomena of drying of colloids in a model sessile drop configuration. Then we describe two important attributes of drying colloid, they are

†‡Contributed equally
*Corresponding author:hisaylama@gmail.com

(a) pattern formation and (b) desiccation cracks. The complete outline for the chapter is as follows:-

- Introduction
- Drying of drop
    — Theoretical background
    — Evaporation driven flow
        * capillary flow
        * inter-facial flow
        * Marangoni flow
- Dried deposit pattern
    — Parameter influencing dried deposit
        * particle shape and size
        * particle surface charge
        * drying condition
        * substrate wettability
        * geometrical confinement
- Desiccation cracks
    — Theoretical background
        * stress and strain
        * strain energy
    — Crack pattern
        * effect of physico-chemical condition
        * impact of mechanical property
        * role of local micro-structure
        * cracks under external field
- Summary

**DRYING OF COLLOIDS**

The phenomenon of drying of colloid mainly comprises: (i) diffusion of vapour from the liquid-air interface to the ambient atmosphere via evaporation induced drying, (ii) advection of the constituents and their accumulation into a particle deposit [3,5]. These phenomena have been widely studied via several standard model experiments. Out of which drying in a droplet configuration provides a lot of theoretical insights into the subject. There is again two standard configurations in the droplet, namely - (a) contact-free drop and (b) sessile drop. The contact-free drop is a spherical drop without a substrate (shown in Fig. 2(a)) while the sessile drop is a drop supported by a substrate (shown in Fig. 2(b)). Here we will discuss only the drying in sessile drops.

†‡Contributed equally
*Corresponding author:hisaylama@gmail.com

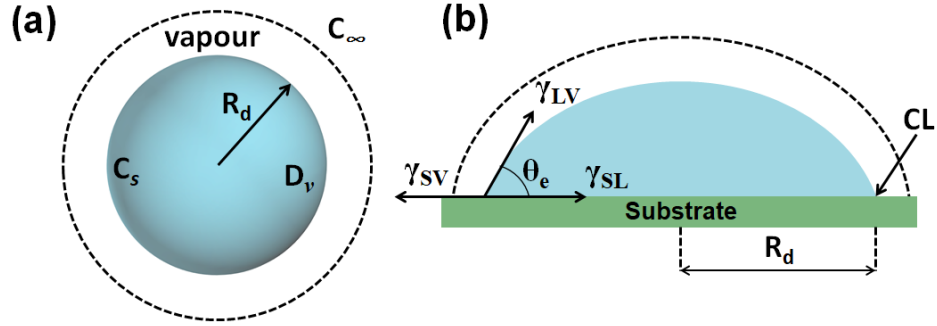

Figure 2: Schematic of drying drops: (a) contact-free spherical drop, (b) drop placed on a substrate (sessile drop) taking the shape of a spherical cap and forming a three-phase contact line (CL). The droplet radius $R_d$, vapour diffusivity $D_v$, concentration of vapour on the droplet surface $C_s$ and the ambient $C_\infty$, equilibrium contact angle $\theta_e$ are pointed. The interfacial surface tensions: $\gamma_{SL}$, $\gamma_{LV}$, $\gamma_{SV}$ for solid-liquid, liquid-vapour and solid-vapour interfaces respectively are also indicated.

## Theoretical Background

When a drop of colloid is placed on a substrate, it looks like a spherical cap and forms three-phase contact line at the edge where the solid, liquid and gas co-exist (depicted in Fig. 2 (b)). This configuration is popularly known as sessile drop. In principle the process of drying is characterized by two time-scales viz. diffusion time scale ($\tau_{diff}$) and evaporation time scales ($\tau_{evap}$). The former is related to the diffusion of vapour into the ambient atmosphere while the latter is the time required to completely evaporate. For the droplets, they are expressed as [5],

$$\tau_{diff} \sim \frac{R_d{}^2}{D_v} \tag{1}$$

$$\tau_{evap} \sim \frac{R_d}{dV_d/dt} \tag{2}$$

where $D_v$ is the diffusion coefficient of vapour in the atmosphere, $R_d$ is radius of a drop (indicated in Fig. 2(a)-(b)) and $dV_d/dt$ is the rate of change in volume of a droplet. The relative magnitude of these timescales dictates the resultant pattern formed exhibited by the dried residue. For example, when $\tau_{diff}/\tau_{evap} \ll 1$, the evaporation is called diffusion limited.

†‡Contributed equally
*Corresponding author:hisaylama@gmail.com

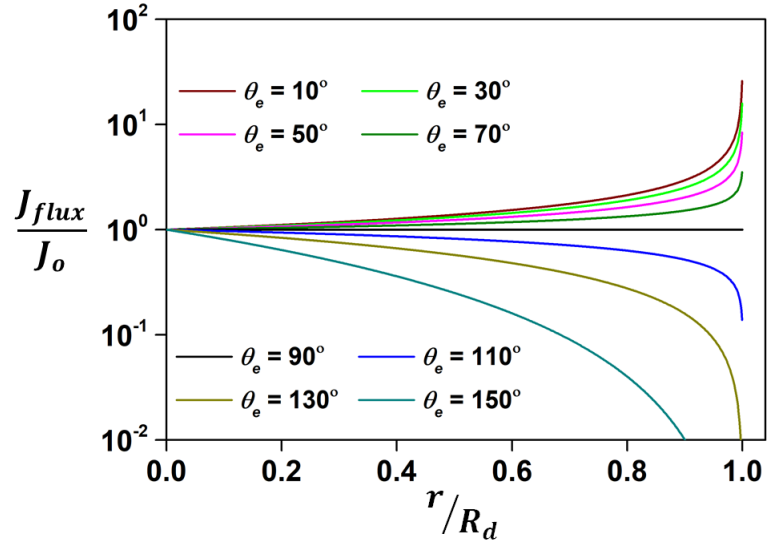

Figure 3: Plot of dimensionless local diffusive evaporative flux ($J_{flux}/J_0$) on the droplet surface as a function of scaled radial distance ($r/R_d$) for various $\theta_e$ (10° to 150°).

For a drop placed in sessile configuration, the droplet first spreads on a substrate then reaches to an equilibrium state and gets pinned with an angle $\theta_e$, at the three-phase contact line. The shape of a drop is determined by $\theta_e$, and the capillary length (defined as $l_c = \gamma_{LV}/\rho g$, where $\gamma_{LV}$ is the liquid-vapour interfacial tension, $\rho$ is the density of the liquid and $g$ is the acceleration due to gravity). The evaporative flux $J_{flux}$ on the surface of a drop is given by [6,7],

$$J_{flux}(r,t) \propto \left(1 - \frac{r}{R_d}\right)^{-\left(\frac{\pi-2\theta(t)}{2\pi-2\theta(t)}\right)} \qquad (3)$$

where $r$ is arbitrary location towards the contact line, $R_d$ is the radius of a sessile drop, $\theta(t)$ is the instantaneous contact angle of a drop on the substrate. The Eq. 1 can be further simplified as,

$$J_{flux}(r,t) = J_0 \left(1 - \frac{r}{R_d}\right)^{-\left(\frac{\pi-2\theta(t)}{2\pi-2\theta(t)}\right)} \qquad (4)$$

where $J_0$ is a proportionality constant. As evident from Eq. 3 and Eq. 4, the evaporative flux $J_{flux}$ is non-uniform across the surface and depends on the equilibrium contact angle ($\theta_e$) of a drop on the substrate. The variation in evaporative flux along the drop surface with different $\theta_e$ are plotted and shown in Fig. 3. This $\theta_e$ is also a characteristic of substrate wettability and depending on their magnitude they are classified as - (i) hydrophilic substrate ($\theta_e < 90°$), (ii) neutrally wetting substrate ($\theta_e = 90°$), (iii) hydrophobic substrate ($\theta_e > 90°$), and (v) super-hydrophobic substrate ($\theta_e > 150°$). From Fig. 3, it is apparent that for a sessile drop with $\theta_e < 90°$ i.e. hydrophilic substrates, $J_{flux}$ is maximum at the contact line and tend to diverge. While at the apex, $J_{flux}$ is minimum and attains a constant value. In contrast, for the hydrophobic or super-hydrophobic substrates i.e. for $\theta_e > 90°$, it's the reverse, $J_{flux}$ is maximum at the apex while the minimum at the edge. Interestingly, when $\theta_e = 90°$, i.e. for a neutral substrate, $J_{flux}$ is constant and is uniform across the complete droplet surface. This non-uniformity in evaporative flux across the droplet surface significantly affects the deposition of comprising colloids and their resultant pattern. For example, when $\theta_e << 90°$ i.e. hydrophilic substrate, the non-uniformity in $J_{flux}$ is such that there is a radial outward flow of fluid together particles resulting in ring-like deposit popularly known as a coffee-ring deposit. The expression for flux from Eq. 3 can be written as [6,7],

$$J_{flux}(r) \propto \frac{1}{\sqrt{(R_d - r)}} \tag{5}$$

The formation of coffee-ring like deposit is a widely studied phenomenon that has important implications in the research of colloidal drying and pattern formation.

Similarly, for contact-free drops the evaporative flux $J_{flux}$, on the surface of a drying colloidal drop can be expressed as [5],

$$J_{flux} = \frac{D_v}{R_d}\left(\frac{C_s - C_\infty}{C_0}\right) \tag{6}$$

where $C_0$ is the concentration of the pure liquid, $C_s$ and $C_\infty$, are the concentration of vapour at the droplet surface and infinity respectively. Unlike the case of a sessile drop, here, the evaporative flux along the air-liquid interface is constant.

**Evaporation Driven Flow**

The formation of a colloidal deposit by drying driven processes involves the local transportation of dispersed particles via the local flow field induced in the dispersion. The flow fields originating from the evaporation induced flows plays a crucial role in the formation of a deposit and their emergence into a distinct pattern. For a colloidal dispersion drying in a sessile configuration, the flow field inside an evaporating drop is linear [21,22], closed loops [21,22], inter-facial or any combination of them as schematically shown in Fig. 4. All of these mentioned flow fields have different origin and are known to be dependent on the drying geometry, motion of contact line and the physicochemical condition of drying. All of these flow fields have a different origin. Here, we briefly discuss few, most important ones that are known to exist for the drying sessile drop - (a) capillary flow, (b) interfacial flow and (c) Marangoni flow.

†‡Contributed equally
*Corresponding author:hisaylama@gmail.com

***(a) Capillary flow*** – The capillary flow is known to be omnipresent in the sessile drops that are drying on hydrophilic substrates. Their origin is attributed to the non-uniform evaporation flux on the droplet surface and is also known to result in "coffee-ring" deposit. In a drying sessile drop, it generates radial outward flow field and transports the particles to the edge via advection. The particle migration velocity due to the capillary flow at any radial location ($r$) can be estimated using the following expression [23],

$$U_c \sim \frac{D^*}{\theta(t)\sqrt{R_d(R_d - r)}} \tag{7}$$

where $D^* = \frac{2\sqrt{(2)}D_v\Delta c}{\pi\rho}$ is the effective diffusivity of the fluid in the droplet with $D_v$ being the diffusion constant for vapour in air, $\Delta c$ is the difference in concentration of liquid-vapour at the droplet surface and the ambient environment and $\rho$ is a density of fluid, $\theta(t)$ is instantaneous three-phase contact angle, $R_d$ is the contact radius of a droplet.

As evident from Eq. 7, the capillary flow velocity is inversely proportional to $\theta(t)$ and vary with the radial distance $r$ within the droplet. This resembles the higher velocity of the fluid at the region close to the droplet contact line and is equivalent to a situation of squeezing. This results in an outward radial flow with the deposition of particles at the edge. Note that towards the tail end of drying, i.e. when $\theta(t) \to 0$, capillary flow velocity ($U_c$) diverges, the flow field is maximum and becomes independent of $r$. As mentioned previously, the strength of this capillary flow greatly depends on the variation in evaporative flux $J_{flux}$, along the drop surface which is further known to depend on the substrate wettability and evaporation rate.

†‡Contributed equally
*Corresponding author:hisaylama@gmail.com

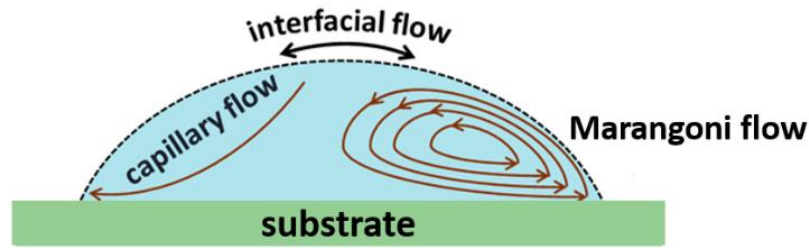

Figure 4: Schematic representation of different flow fields inside the evaporating droplet. They are originated via capillary flow, Marangoni flow and the interfacial flow, arrows point the direction of fluid flow.

***(b) Interfacial flow*** –The interfacial flow is another important mechanism of advection that are also common in the drying colloidal systems. They typically originate from the gravity-induced deformation of a drop comprising of colloidal particles [24]. In the drying dispersion particularly the sessile drop, during the evaporation, the particles are adsorbed to the air-liquid interface and get trapped. These trapped particles at the interface require high energy for their detachment and therefore prefers to remain there instead of getting detached. Now to reduce the interfacial energy, the drying droplet chooses an alternate route where the adsorbed particles transport through the air-liquid interface. The particles at the air-liquid interface are known to move from the regions of higher surface curvature (apex) to that of lower (edge) [25]. Therefore, this process results in an accumulation of the particles at the droplet edge. Note that these flow mechanisms are prevalent for the substrates where the $\theta \geq 90°$. The strength of this inter-facial flow driven accumulation of particles depends upon the mean and deviatoric curvatures of the air-liquid interface [25].

***(c) Marangoni flow*** – The Marangoni flows inside a drying drop is generated due to the gradient in the surface tension along the drop surface. Typically, the direction of the Marangoni flow is from the region of lower surface tension region to that with higher surface tension. This flow can be induced either by generating the temperature gradient along the drop surface (often referred as thermal Marangoni effect [22,26]) or by adding the surface-active solutes such as polymers or surfactant (referred to as the solutal Marangoni effect [27–29]). For a sessile drop evaporating at a fixed temperature, the gradient in surface tension is developed as a result of differential cooling along the interface and originates from non-uniform evaporative flux along drop surface. The temperature at the apex of a drying sessile drop is lesser than at the contact line. This local variation in temperature depends on the relative magnitude of the thermal conductivity of a substrate and the dispersion [30]. As the surface tension is inversely proportional to the temperature, therefore there is a gradient in surface tension along the droplet surface such that magnitude of surface tension is highest at apex and lowest at the contact line. Therefore, there is an inward flow field followed by the advection of particles in the bulk region of a drop. The velocity of this flow field is given by [26],

$$U_{Ma} = \frac{1}{32} \frac{\beta \theta(t)^2 \Delta T}{\eta} \tag{8}$$

where, $\beta$ = gradient in the surface tension,
$\theta(t)$ = instantaneous contact angle of the drop,
$\eta$ = dynamic viscosity of the fluid and
$\Delta T$ = difference between the maximum and minimum temperature in the drop.

†‡Contributed equally
*Corresponding author:hisaylama@gmail.com

As evident from the Eq. 8, the velocity of fluid that is driven by Marangoni flow is directly proportional to $\beta$ and $\Delta T$ while inversely proportional to $\eta$. Thus increasing $\Delta T$ or $\beta$, can enhance the Marangoni flow driven deposition of particles.

Similarly, to the thermal Marangoni flow, an addition of surface-active agents (or surfactants) to an evaporating droplet can induce solutal Marangoni convection of the particles [31]. In such cases, the outward capillary flow first advects the surfactant molecule towards the contact line which in turn lowers the surface tension at that region compared to the droplet apex. This sets up the gradient in surface tension with being minimum at the edge and maximum at the contact line. Therefore, this generates an inward flow of fluid together with the constituting particles towards the droplet interior [32].

Hence, for a sessile drop comprising of colloidal particles, the interplay between these flow fields can alter the resultant dried. For example, when a sessile drop comprising particles drying in sessile configuration, the dominance of capillary flow result in coffee-ring deposit while that of Marangoni flow results in its suppression [22].

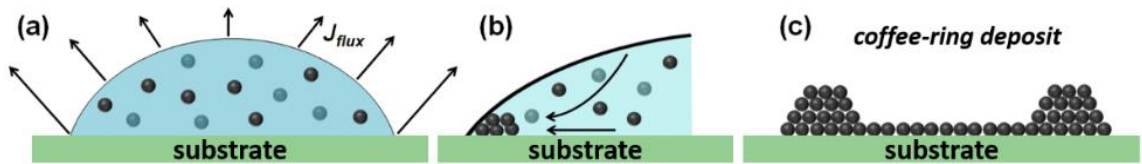

Figure 5: Schematic representation of (a) the side view of an evaporating sessile drop containing dispersed particles with the arrows indicating the local evaporative flux ($J_{flux}$), (b) transportation of particles toward the contact line denoted by the arrows and (c) the side view of the final dried particulate deposit showing the accumulation of most particles at the drop edge after solvent evaporation.

†‡Contributed equally
*Corresponding author:hisaylama@gmail.com

## DRIED DEPOSIT PATTERNS

As mentioned previously, the drop of colloid dried in a sessile configuration often leaves a ring-like deposit named *coffee-ring*. Deegan et al. presented the first quantitative analysis and explains the formation of such ring-like deposit [6,7]. The origin of coffee-ring deposit was ascribed to the pinning of a contact line and the non-uniform evaporative flux ($J_{flux}$) that sets across the droplet surface. The magnitude of $J_{flux}$ varies such that their value is maximum at the contact line and minimum at the apex (depicted in Fig. 5 (a)).

This non-uniformity in $J_{flux}$ generates an outward radial flow named capillary flow from the bulk of a drop towards the contact line (indicated in the schematic shown in Fig. 5 (b)). The capillary flow compensates for the higher loss of fluid from the contact line and transports the dispersed particles towards the droplet edge. On complete drying, the accumulated particles manifest into a thick ring-like deposit that comprises of closely packed particles as shown in Fig. 5 (c).

The thickness, width and radius of this ring-like deposit constituting colloidal particles are known to depend on several physical parameters including initial volume fraction of the dispersion, wettability and the drying conditions [7]. The deposition of the colloidal particles is a result of their accumulation during drying via evaporation driven flows. In the subsequent section, we will discuss them in detail.

### Parameters Influencing Dried Deposit

The dried deposit emerging into distinct patterns are controlled by various intrinsic (viz. shape, size and charge of the constituting particles) and extrinsic (viz. substrate wettability and drying condition) parameters. In this section, we will briefly describe the role of these different parameters.

*(a) Particle size and shape* – The size of the particle is found to play a pivotal role in controlling the deposit patterns. The colloid evaporating in sessile configuration, containing smaller size particle (diameter $\approx$ 2 μm) predominantly forms coffee-ring like deposits. While that comprising of bigger size particles (diameter $\approx$ 20μm) forms multiple ring-like deposit (shown in Fig. 6(a)) [33]. The variation in the size of the suspended particles alters the flow field inside the evaporating drop and are known to originate from the reverse capillary motion of fluid [33]. Apart from the size of a constituents particles their shape has also been shown to alter the deposit pattern. For example, Yunker et al. has shown that for a sessile drop containing polystyrene sphere, the drying yields coffee-ring deposit (shown in Fig. 6(b)), while that with an aspect ratio (AR) greater than 1.5, uniform deposit occurs (Fig. 6(c)) [34]. The formation of the uniform deposit is attributed to the adsorption of the ellipsoids at the water-air interface that prevents their deposition at the contact line.

†‡Contributed equally
*Corresponding author:hisaylama@gmail.com

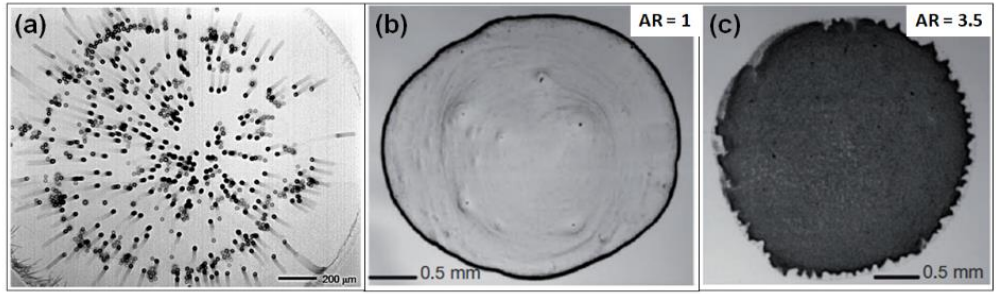

Figure 6: Optical microscopy images of (a) drying droplet showing the reverse capillary motion of the dispersed particles (10µm polystyrene microspheres) [33]. Optical microscopy images of a dried droplet (a) with polystyrene micro-spheres shows coffee-ring deposit [34] while (c) with polystyrene ellipsoids exhibit uniform deposit [34].

*(b) Particle surface charge* – Besides size and shape of the particles, the surface charge of the constituting particles is also found to play a crucial role in the process of consolidation of colloidal particles. In principle, colloidal particles dispersed in a continuous medium can be assumed as a dispersed dielectric sphere that experiences a repulsive electrostatic force and attractive van der Waals (together called DLVO force). The surface charge $\Sigma_p$ on these particles are given by [35],

$$\Sigma_p = -\frac{F\Delta n}{m_p S_p} \tag{9}$$

where, $F$ is the Faradays constant (96500 C/mol), $\Delta n$ is the number of ions adsorbed onto the surface of particles, $m_p$ is the mass of the particles (in kg), and $S_p$ is the specific surface area of a particle (in $\mathrm{m^2/kg}$). The $\Sigma_p$ on particles is modulated by controlling the number of $H^+$ or $OH^-$ adsorbed ions, i.e. by changing their pH. Similarly, the number of ions that are accumulated on the substrate and the net interaction force exerted on them can also be modulated by varying pH of a solvent. Bhardwaj et al. have shown that for a dispersion comprising of Titania particles, dried in a sessile configuration, the variation in pH results in distinct dried patterns as shown in Fig. 7 [35]. At pH < 5 and pH > 9, the dried deposit is coffee-ring like, while in a range 5 < pH < 9 the deposit is uniform. The variation in deposit profile at different pH is attributed to the relative strength of interactions forces between the particles and the particle-substrate. A similar observation is reported by Dugyala et al. for the dispersion of non-spherical hematite ellipsoids as shown in Fig. 8 [36]. As evident from Fig. 8, a coffee-ring deposit is obtained when an aqueous drop at highly acidic (pH < 5) or highly basic pH (> 9) is dried, however at an intermediate pH the deposit is uniform.

†‡Contributed equally
*Corresponding author:hisaylama@gmail.com

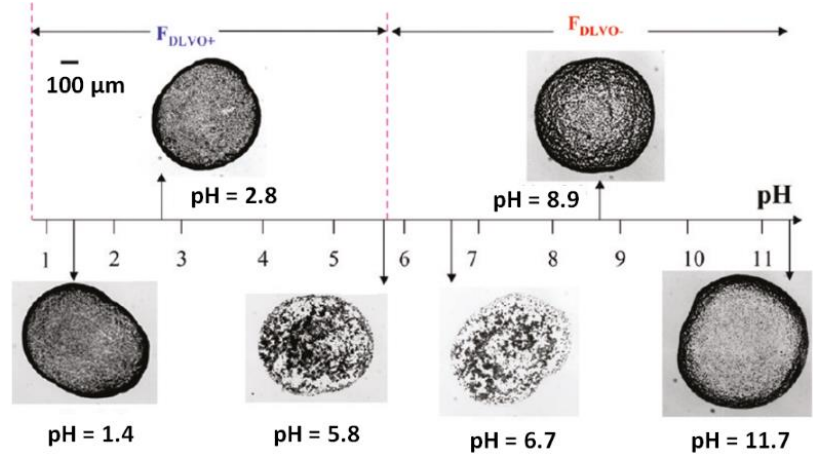

Figure 7: Images of dried droplets showing the variation in their dried patterns as a function of pH of a drying dispersion [35]. All of these drops contain Titania nano particles of diameter = 25nm possess initial concentration = 2% (v/v) and were dried at temperature 25°C.

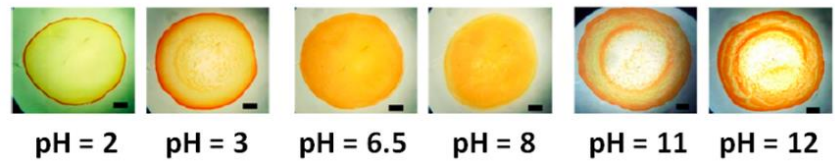

Figure 8: Images of the dried droplets containing hematite ellipsoids of A.R= 4 (with length/breadth = 200nm/50nm with varying pH shows the variation in the deposit pattern. All the drops were obtained at temperature= 25°C and RH=40%). The scale bar in each image corresponds to 500 μm) [36].

*(c) Drying condition* – The environmental conditions under which the drying experiments are performed can potentially alter the nature of the deposit patterns [37–39]. The evaporation rate and flow field significantly vary on changing the ambient condition particularly the substrate temperature ($T_{sub}$) and relative humidity (RH). They directly affect the rate of diffusion of liquid-vapour from the air-water interface to the ambient atmosphere and the mobility of the particles inside the dispersion. For a sessile drop of colloids comprising of polystyrene micro-spheres and drying at different temperatures, Li et al. have shown that the deposit can undergo the transition from ring-like deposit to a uniform deposit (shown in Fig. 9(a) - (b)). For a sessile drop drying at a temperature $T_{sub} \approx$ 25°C, the deposition is mostly governed by outward capillary that results in the deposition of particles at the edge and result in ring-like deposit as shown in Fig. 9(a). While the deposit drying at a greater temperature ($T_{sub} > 70°C$), the air-water interface of a drop is found to descend at a significantly higher rate. This result in the capture of particles at an interface and hinders the capillary-flow driven deposition of particles at the edge (prevalent at $T_{sub} \approx 25°C$). Finally, a uniform deposit is obtained as shown in Fig. 9(b).

†‡Contributed equally
*Corresponding author:hisaylama@gmail.com

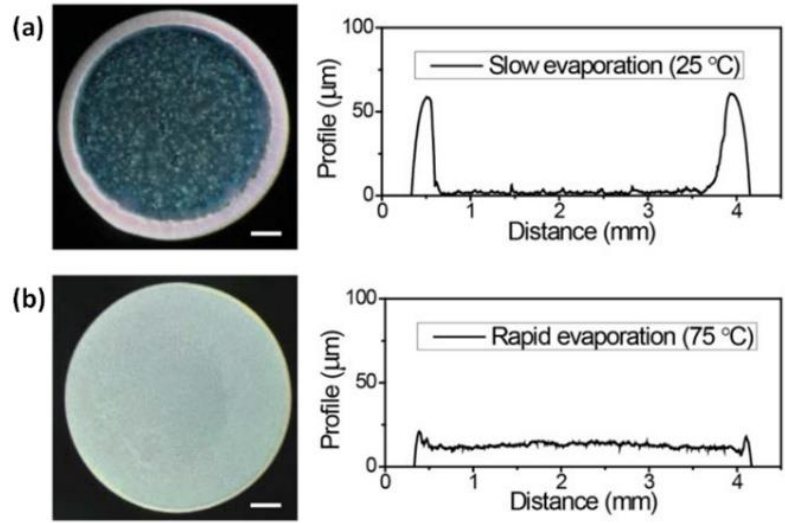

Figure 9: Optical microscopy images of the dried sessile drops comprising polystyrene microspheres, obtained - (a) at room temperature $T_{sub} \approx 25°C$) and (b) at an elevated temperature $T_{sub} \approx 25°C$ ) with constant relative humidity, RH $\approx 50\%$. The spatial variation in their height profiles are depicted in the right panel [37].

Besides, the interface capture of particles there is thermal driven Marangoni flow that can modify both the flow field and the deposit pattern [40]. Other popular schemes that are also exploited to alter the flow field and the deposited pattern are drying in the presence of (a) electric field [41] and (b) the acoustic waves [42].

*(d) Substrate wettability* – As evident from Eq. 3 and Eq. 4, the evaporative flux Jflux can be seen to vary with the equilibrium contact angle $\theta_e$. Moreover, the uniformity and non-uniformity in $J_{flux}$ across the air-fluid interface also depend on the magnitude of $\theta_e$. This induces a notable change in the flow field that is generated during drying. For example, a sessile drop drying on a hydrophilic substrate, i.e. with $\theta_e < 30°$, the dominant mode of transport is outward capillary flow. While, when $\theta_e > 30°$, the flow field inside the drop are reported to be the closed loops (depicted in Fig. 3). They are generated via thermal Marangoni flow [21,22]. Conventionally, the drying of a colloidal dispersion is mostly studied on the hydrophilic substrates i.e. with $\theta_e < 90°$. The nature of the deposit is often ring-like shown in Fig. 10 (a1) - (c1). The deposition of constituting particles at the contact line via evaporation induced flow decreases with the increase in $\theta_e$. For a colloid drying on a hydrophobic substrate, it often yields a uniform deposit as shown in Fig. 10 (d1). The possibility of generating the structured surfaces has made it possible to investigate the evaporation induced drying studies on the hydrophobic and super-hydrophobic surfaces i.e. with $\theta_e > 90°$ [43,44]. They often form multiple rings of particulate deposits. Further, this investigation has been extended as the function of surface roughness. They are found to significantly affect the formation of the deposit patterns and influence the spatial distribution of particles in the dried deposit [45].

†‡Contributed equally

*Corresponding author:hisaylama@gmail.com

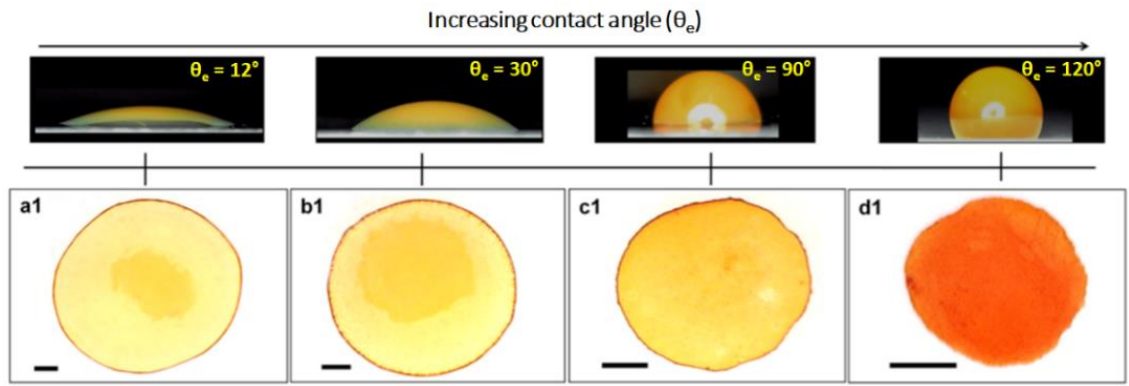

Figure 10: Optical microscopy images of the dried sessile drops containing hematite ellipsoids of AR= 4.3 (maintained at pH ≈ 2) on the substrates with various $\theta_e = [12°, 120°]$. All the deposit are obtained at temperature= 25°C and RH= 50%. The scale bar in each image corresponds to 500 μm [24].

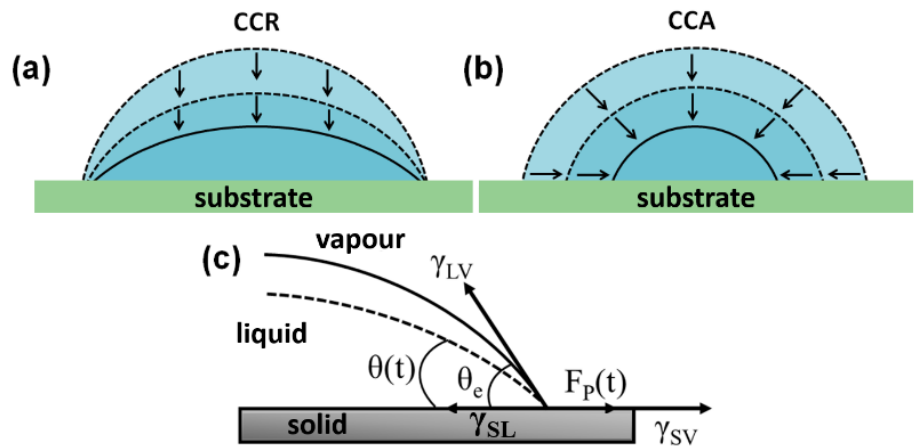

Figure 11: Schematic shows the contact line dynamics of an evaporating droplet. (a) Constant contact radius (CCR) mode where the diameter of the drop remains constant during drying and (b) Constant contact angle (CCA) mode where the contact angle remains the same and diameter of the drop decreases during drying. The arrows indicate the direction of receding of a droplet surface. (c) Schematic showing the force balance along the pinned three-phase contact line.

†‡Contributed equally

*Corresponding author:hisaylama@gmail.com

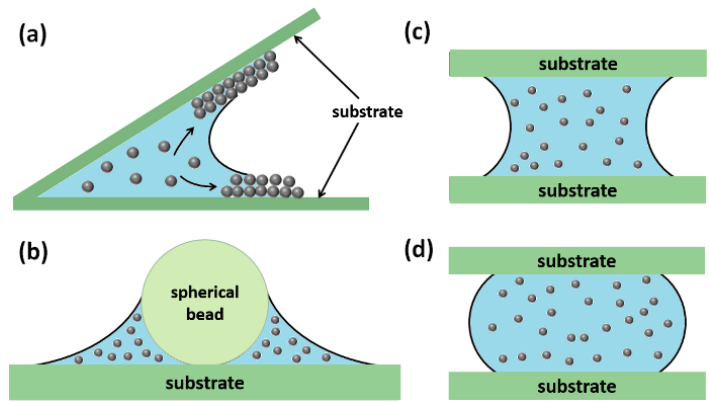

Figure 12: Schematic of different drying configurations - (a) wedge geometry, (b) sphere-on-plate, (c) confined between two hydrophilic plates and (d) hydrophobic substrates.

**(e) Effect of geometrical confinement** – Drying the colloidal dispersion under geometrical confinement expected to gain control over the evaporation process by manipulating the contact line dynamics. Therefore, the investigation of contact line dynamics during the process of drying is very essential to develop the methodology for controlling the evaporative induced pattern formation [46,47].

In a sessile drop drying experiments, the evaporation occurs mainly in two different modes: (i) constant contact radius (CCR) mode and (ii) constant contact angle (CCA) mode. In CCR mode of evaporation, the drop evaporates such that the droplet is pinned and the contact area is constant during drying as depicted in Fig. 11(a). While in CCA mode of evaporation, the contact angle is constant and the contact radius decreases during the drying as shown in Fig. 11(b). The evaporation process can be more complex when the two evaporation modes occur one after the other or occur simultaneously [48]. The evaporation with a combination of two modes (CCR + CCA) are typically referred as mixed-mode of drying and it results in "stick-slip" motion of the contact line.

Typically, for a drop evaporating on a hydrophilic substrate i.e. with $\theta_e < 90°$, the contact line is initially pinned or "stick" to the substrate at an initial drying stage i.e. $\theta(t) > \theta_R$, ($\theta(t)$ is the instantaneous CA and $\theta_R$ is the receding CA). The contact line of a drop de-pin or "slip" when $\theta(t) \leq \theta_R$. For a pinned contact line, the force balance at the contact line yields (schematic shown in Fig. 11(c)),

$$\mathrm{F}_p(t) = \gamma_{LV} cos\theta(t) + \gamma_{SL} - \gamma_{SV} \tag{10}$$

where $F_p(t)$ is an instantaneous pinning force per unit length, $\gamma_{LV}$, $\gamma_{SL}$, and $\gamma_{SV}$, are interfacial tensions of the liquid-vapour, solid-liquid and solid-vapour interfaces respectively. The droplet is known to de-pin when the pinning force attains its maximum possible value and $\theta(t) \approx \theta_R$.

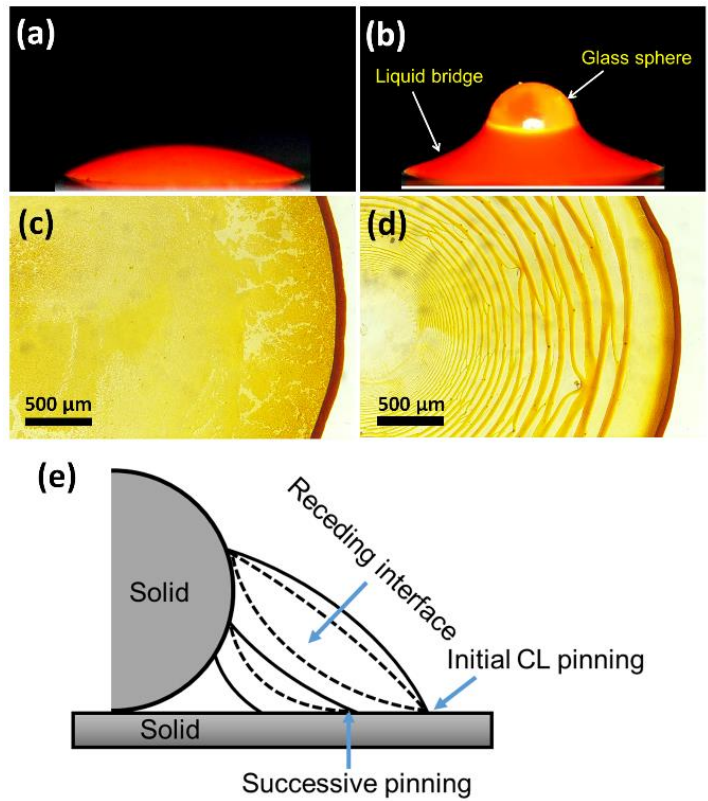

Figure 13: The dispersion of hematite ellipsoids (AR= 5.18) maintained at pH ≈ 2, drying in (a) sessile and (b) sphere-on-plate configuration. Optical microscopy images of the dried deposits obtained for suspension dried in (c) sessile and (d) sphere-on-plate configuration. (e) Schematic depicts the motion of a contact line for the dispersion drying in the sphere-on-plate configuration [49].

The "stick-slip" motion of contact line during the drying can either be discrete or continuous. In a discrete stick-slip motion, the pinning force along the contact line is assumed to be same everywhere and the motion of the contact line is symmetric. This directly implies that during drying the entire contact line de-pins and subsequently gets pinned to a new location simultaneously. While during the continuous stick-slip motion of a drop, the entire contact line does not de-pins symmetrically. The interplay of contact line dynamics together with the geometrical confinement has been exploited to generate various deposit pattern. Some examples of geometrically confined drying are drying (i) inside a wedge geometry, (ii) sphere-on-plate or (iii) confined between two parallel plates as schematically shown in Fig. 12(a) - (d). The drying of colloid inside the wedge-like geometry has been exploited to create a mono-layer assembly of colloidal particles via slow evaporation of a solvent. Similarly, the colloids dried in sphere-on-plate configuration or confined between two parallel plates are used to generate the directed self-assembly of colloidal particles resulting in novel deposit patterns. The drying of colloidal dispersion, particularly in sphere-on-plate configuration results in the formation of concentric multi-ring-like deposit as shown in Fig. 13. Such a concentric multi-ring like deposits are formed due to discrete stick-slip motion of the contact line. In contrast, when the pinning force varies locally along the contact line, an asymmetric motion of the contact line was observed. In those cases, the contact line remains pinned in some regions and simultaneously de-pins in other regions. This results in continuous stick-slip motion of the contact line. Such continuous stick-slip motion was observed by Mondal et al. for colloids drying between two parallel plates. The resultant deposits are found to be spiral ring-like patterns as shown in Fig. 14 [50]. It is worth mentioning that the formation of spiral deposit patterns is not affected by the direction of receding of contact line i.e. receding in clockwise direction or an anticlockwise direction or a combination of both.

†‡Contributed equally
*Corresponding author:hisaylama@gmail.com

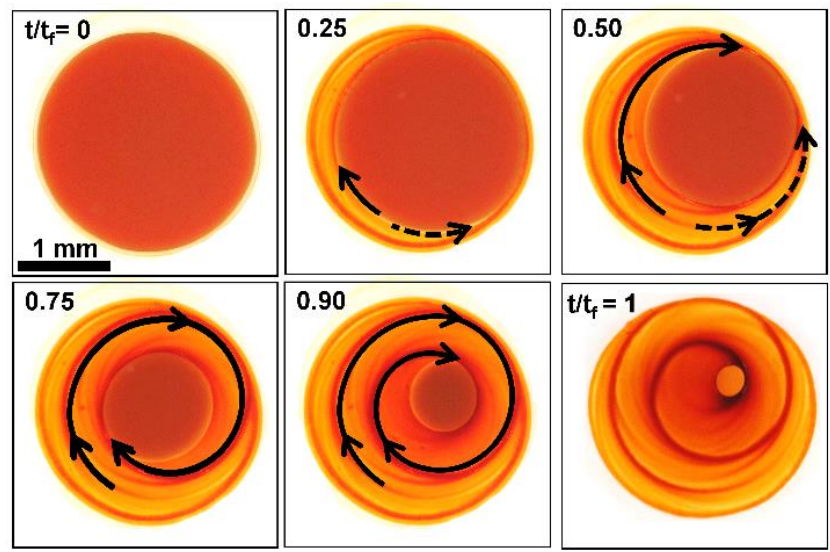

Figure 14: Time-lapse optical microscopy images of the drying dispersion comprising of hematite ellipsoids that are confined between two hydrophilic glass plates. The receding motion of a contact line and their de-pinning are indicated by an arrow. The de-pinning of the contact line, their propagation in clockwise and anticlockwise directions, and the resultant spiral deposit are all depicted [50]. The instantaneous and total drying time is referred by $t$ and $t_f$ respectively.

## DESICCATION CRACKS

As previously mentioned, the drying of colloidal dispersion yields a consolidated solid deposit comprising of closely packed particles [3,4]. For example, *coffee-ring* like deposit was obtained for a colloid dried in a sessile configuration. The dried residue of colloid often possesses a structural defect named crack Fig. 15(a). They are popularly known as "*desiccation cracks*". Besides desiccation cracks, several other structural defects viz. buckling, debonding and warping (shown in Fig. 15(a)-(c)) are also frequently observed in the dried colloidal deposit but here we will only discuss the former. In the laboratory, all of them are studied via model experiment i.e. isotropic or directional drying experiments as mentioned in the introduction section.

†‡Contributed equally
*Corresponding author:hisaylama@gmail.com

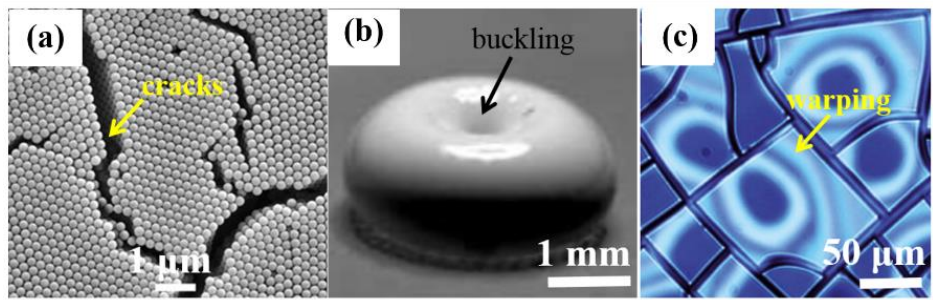

Figure 15: (a) Cracks in the colloidal deposit comprising of poly-methyl methaacrylate (PMMA) [51] (b) Colloid deposit comprising of 100nm latex particles shows buckling at the centre [52], (c) Silica (Ludox©dispersion) particulate film displays the warping as pointed by an arrow [53].

## Theoretical Background

The nucleation of cracks in the solid colloidal deposit occurs in several stages [3,4]. First, the particles in the colloid drying on a substrate consolidate, their physical state transforms from the state of a liquid dispersion to a gel-like and then finally into a solid deposit (shown in Fig. 16(a)-(c)). The loss of fluid in the gel-like state tend to shrink the top layer of gel while the adhering substrate obeys no-slip condition and resists the shrinkage. The process develops frustration and generates strain energy in the deposit. When the accumulated strain energy exceeds its optimum value, the colloidal deposit fails and crack nucleates. Please note that the generation of strain energy in the deposit also implies the modulation of associated tensile stress and strain. The stress and strain are important variables that determine the nature of cracks. The physics of desiccation cracks are typically explained by citing its analogy with the cracks formed in brittle solid. However, the origin of cracks forming in brittle solid material and the colloidal deposit is different. For solid material the crack nucleates from the pre-existing flaw upon an application of tensile stress. In contrast, cracks in the colloidal deposit nucleates from the void between the particles due to the release of intrinsically developed stress. In addition, the nucleation of a crack in the colloidal deposit implicates the physical separation of particles that were initially bonded together Fig. 15(a).

†‡Contributed equally
*Corresponding author:hisaylama@gmail.com

***Stress and strain*** – The stress and strain in a drying colloidal deposit are important measurable variables that facilitate the estimation of associated strain energy. As mentioned previously, they are intrinsically developed via drying induced shrinkage and their magnitudes are also continuously modulated. The stress and strain developed in the deposit can also be obtained theoretically by exploiting the continuum mechanics framework known for the poroelastic materials. This formalism was first derived by M.A Biot in 1941, it is popularly known as "*theory of poroelasticity*" [54]. This theory is a conjunction of fluid mechanics and structural mechanics. Now, assuming the drying gel-like colloidal deposit to be the isotropic linear poroelastic material, the total stress exerted on the deposit can be expressed in index notation as [54–56],

$$\sigma_{ij} = \tilde{\sigma}_{ij} - \alpha \delta_{ij} \prod \tag{11}$$

where, $i, j$ are indices, $\sigma_{ij}$ and $\tilde{\sigma}_{ij}$ is the total stress and the hydrostatic stress exerted on the particle deposit, $\prod$ is the pressure exerted on the void between the particles, $\alpha$ is the Biot's coefficient. The symbol $\delta_{ij}$ is a Kornecker delta with $\delta_{ij} = 0$ for $i \neq j$ and $\delta_{ij} = 1$ for $i = j$. In Eq. 11, the first term represents the stress on the particle network while the second term resembles the contribution of pressure exerted on the void between the particles.

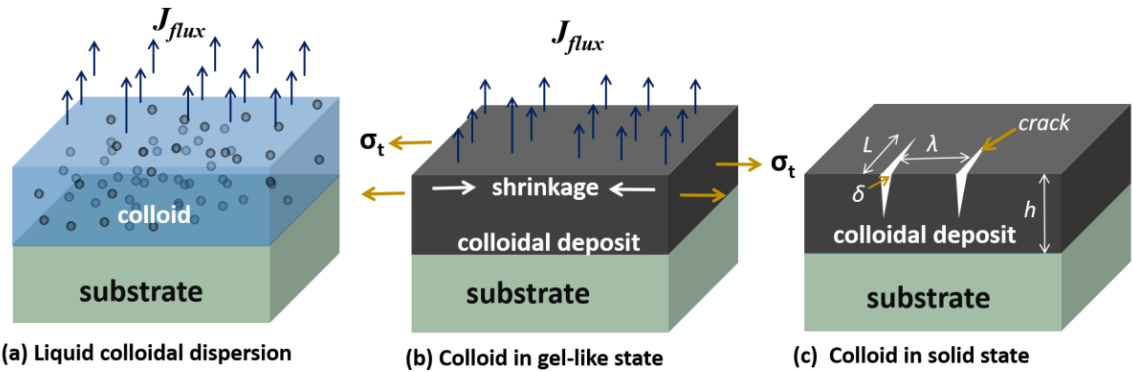

Figure 16: Schematic shows various stages of drying that results in a solid particle deposit. (a) The initial stage of drying with where the colloid is in the liquid state. The $J_{flux}$ is an evaporative flux. (b) Colloids in a gel state due to the loss of fluid, and causing the shrinkage of a deposit. The tensile stress $\sigma_t$ in the deposit that is generated during drying is indicated by an arrow. (c) Solid particle deposit with cracks. The crack length $L$, crack width $\delta$, crack spacing $\lambda$ and thickness $h$ are pointed.

The strain in the deposit can be related to the stress by Biot's constitutive relation. The constitutive relation in an index notation is expressed as [55]:

$$\epsilon_{ij} = \frac{1}{E_f}\big[(1+\nu_f)\sigma_{ij} - \nu_f\sigma_{ll}\big] - \frac{\alpha\prod}{3K}\delta_{ij} \tag{12}$$

where, $i, j, l = $ indices,

$\epsilon_{ij}$ is strain tensor,

$\nu_f$ and $E_f$ are the Poison's ratio and Young's modulus of a dried colloidal deposit,

$\prod$ is the pressure exerted on the void between the particles,

$K\left(=\frac{E_f}{3(1-2\nu_f)}\right)$ is bulk modulus,

$\delta_{ij}$ is Kornecker's delta function,

The value of $\alpha$ in Eq. 11 and Eq. 12 are typically considered to be *one* for most practical purposes [57].

***Strain energy*** – The strain energy per unit volume accumulated in the colloidal deposit can be expressed in term of stress and strain [4], i.e.

$$U_{so} \sim \sigma_{ij}\epsilon_{ij} \tag{13}$$

where, $\sigma_{ij}$ and $\epsilon_{ij}$ are the stress and strain in the index notation. This strain energy per unit volume further can be expressed in terms of free energy per unit volume ($U_F$) associated with the deposit, given by [58],

$$U_F = U_{so} + U_{ex} \tag{14}$$

where $U_{so}$ represents the magnitude of strain energy per unit volume stored in the deposit and $U_{ex}$ resembles the energy per unit volume required to create new surfaces via crack. The quantity $U_{ex}$ is related to surface-free energy by,

$$U_{ex} = 2\Gamma L \tag{15}$$

where $\Gamma$ is the associated surface free energy and $L$ is the length of a surface to be created after crack.

Now as the nucleation of cracks in a colloidal deposit resembles the minimization of free energy. Mathematically, it implies,

$$\frac{dU_F}{dL}\bigg|_c = 0 \Rightarrow \frac{dU_{so}}{dL}\bigg|_c + \frac{dU_{ex}}{dL}\bigg|_c = 0 \tag{16}$$

$$or, \frac{dU_{so}}{dL}\bigg|_c = \frac{dU_{ex}}{dL}\bigg|_c = 2\Gamma \tag{17}$$

†‡Contributed equally
*Corresponding author:hisaylama@gmail.com

where the subscript '$c$' resembles the critical. The Eq. 17 implies that the crack propagates when the stored strain energy per unit volume per unit length is equal to the energy required for creating two new surfaces. Therefore, the condition of extremum for $U_F$ gives the optimum/ critical strain energy per unit volume for creating new surfaces via crack. This imply Griffith's criterion for fracture [58]. The L.H.S in Eq. 17 is often called energy release rate or fracture energy and represented by $G_c$ [58]. Similarly the quantity $\frac{dU_{so}}{dL} \equiv \frac{d(\sigma_{ij}\,\varepsilon_{ij})}{dL}$, is conventionally called as energy release rate and symbolically represented as $-G$ where the negative sign with $G$ indicates the net decrease in their magnitude via crack. It should be noted that the quantity $G_c$ has a constant value for a given material. For the drying porous system, its value is reported to be in a range from 0.1 J/m$^2$ to 1 J/m$^2$ [4].

For a thin elastic plate of thickness $h$ and Young's modulus $E_f$ , under tensile stress ($\sigma_t$), the energy release rate $G$ associated with single crack is estimated to be [4],

$$G = \frac{\pi L \sigma_t^{\,2}}{E_f} \tag{18}$$

where $L$ is the crack length as indicated in Fig. 16(c). It is worth mentioning that by using equality for energy release rates expressed in Eq. 18, we can also obtain critical stress $\sigma_c$. The critical stress is defined as the minimum tensile stress that a deposit should acquire to fail [59].

**Crack Pattern**

The desiccation cracks in the colloidal deposit often self-organize and emerge into a distinct pattern [60]. In numerous occasions, the cracks are frequently observed to be regular [61–65] or self-similar [66]. A plethora of patterns are known for the desiccating colloidal deposit, they are, polygonal [60], linear [61,67], wavy [68], spiral [69], radial [70] and circular [70] as shown in Fig. 17(a)-(f). In most cases, cracks pattern in the colloidal deposit are characterized by several physical quantities, such as − (a) spacing between two cracks ($\lambda$), (b) gap between cracks ($\delta$), (c) thickness of cracking material ($h$), (d) length of the crack ($L$), (e) the size of the flaw (f) area of cracking domain ($A_d$) and (g) desiccation time. These parameters allows us to quantify the property of a cracks pattern. For example, Komatsu et al. [63] and Allain et al. [61] reported that for a parallel cracks in the colloidal deposit $\lambda \sim h^{2/3}$, while Goehring et al. reported $\delta \sim L^{1/2}$ [71]. Interestingly, a similar type of emergence of crack into distinct patterns are also common in nature, e.g. polygonal cracks in dried clay [11,60,72], interconnected cracks in old paintings [10] Fig.

†‡Contributed equally
*Corresponding author:hisaylama@gmail.com

18. The analogous properties of desiccation cracks exhibited by the colloidal deposit and nature have urged researchers to investigate for a universality theory relating them [11].

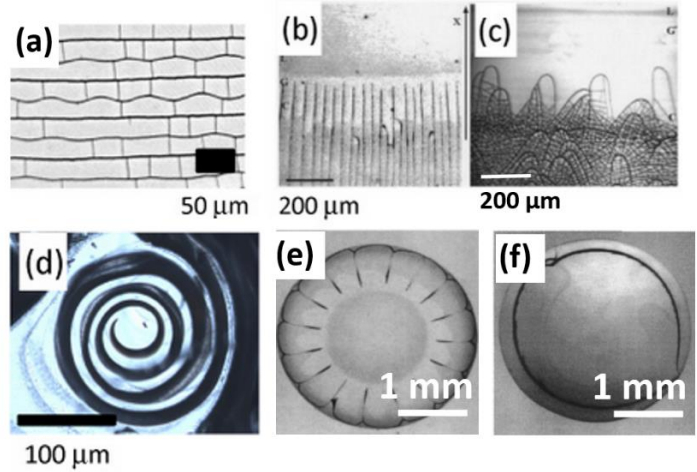

Figure 17: Various morphologies of cracks in the colloidal deposit - (a) wavy cracks [68], (b) linear cracks [67], (c) arch like crack [67], (d) spiral crack [69], (e) radial crack [70] (f) circular cracks [70].

The emergence of cracks into various distinct patterns are known to be affected by numerous parameters. Few of them that have a significant impact is –

- physicochemical condition of drying [73,74],
- mechanical property of colloidal deposit [59,73,75],
- local micro-structure constituting particles [74,76]

The role of these aforementioned parameters on the nucleation of crack and their manifestation into different pattern are described in the subsequent section.

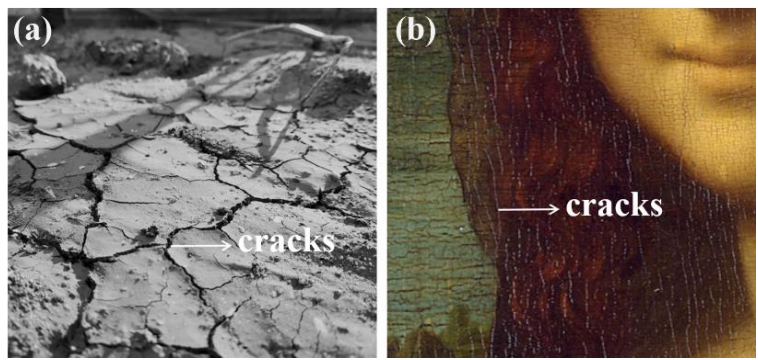

Figure 18: The picture of (a) dried mud with cracks (taken at IIT Madras) and (b) the celebrated painting exhibiting linear cracks are shown [Taken from:htttp:www.fast.u-psud.fr/pauchard/].

†‡Contributed equally
*Corresponding author:hisaylama@gmail.com

***Physicochemical condition*** – Various physicochemical parameters namely salinity [70] and pH of colloidal suspension [77], relative humidity [73] and temperature [78,79] are reported to affect the desiccation crack pattern. In fact, for a specific physicochemical condition, the crack pattern is observed to be distinct. These physicochemical parameters directly affect the drying kinetics due to which the process of consolidation of the colloidal particles is either accelerated or decelerated. This further influences the magnitude of strain energy that develops in the dried deposit. The dependence of physicochemical aspects on desiccation cracks is typically characterized by the characteristic time scales associated with desiccation. They are, (i) gelation time ($'t_G'$) i.e. the time required for a drying dispersion to transform into a semisolid and (ii) desiccation time ($'t_D'$) i.e. the time required for dispersion to completely transform into a solid. For example, Pauchard et al. [70] have demonstrated for a colloid drying in a sessile geometry the addition of salt fastens the gelation time such that $t_G \ll t_D$, with all other drying conditions remains the same (shown in Fig. 19), that eventually vary the resultant morphology of crack. Interestingly, with a varying ionic strength in the dispersion, the crack pattern exhibited by the deposit are distinct (radial or disordered or circular) as shown in Fig. 19. Similar manipulation of $t_G$ is also achieved by increasing or decreasing the relative humidity (RH) of the drying environment. Dauphiné et al. [73] has reported that $t_G \ll t_D$ for the colloid drying in sessile drop configuration. They report that increasing the RH of the drying environment results in the suppression of cracks as shown in Fig. 20. Besides the aforementioned factor, the temperature at which colloid dry, decrease or increase the desiccation time $t_D$ and vary the resultant crack pattern.

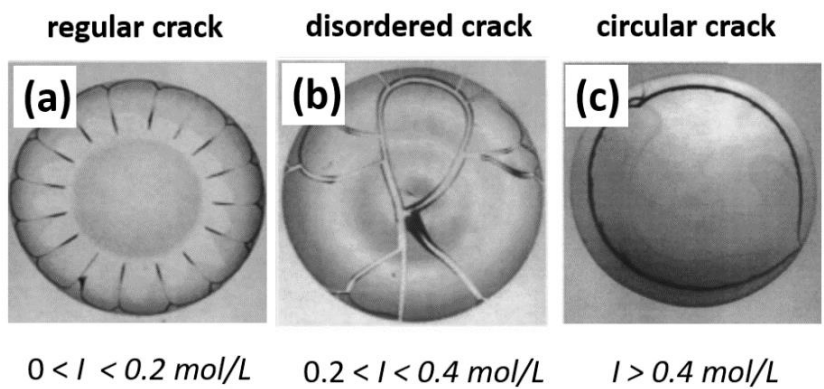

| regular crack | disordered crack | circular crack |
|:---:|:---:|:---:|
| **(a)** | **(b)** | **(c)** |
| *0 < I < 0.2 mol/L* | *0.2 < I < 0.4 mol/L* | *I > 0.4 mol/L* |

Figure 19: Optical microscopy images of dried sessile drops comprising of silica particles (diameter ~ 15 nm, initial volume fraction ~ 0.2) with different ionic strength $I$ (mol/L) and evaporating in an ambient environment. The dried deposit (a) with $0 < I < 0.2$ mol/L exhibits regular radial crack, (b) $0.2 < I < 0.4$ mol/L show disordered crack and (d) for $I > 0.4$ mol/L cracks are circular. The drop base diameter is around 3 mm [70].

†‡Contributed equally
*Corresponding author:hisaylama@gmail.com

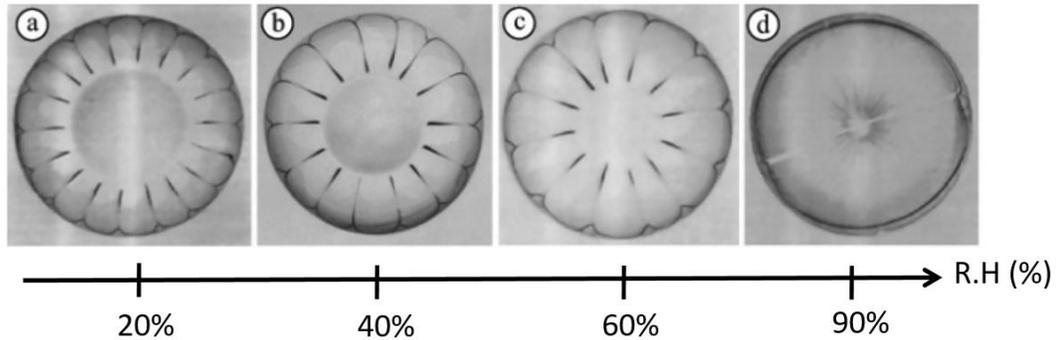

Figure 20: Optical microscopy images of the dried drops comprising of colloids (Ludox® HS-40, $\theta_e$ = 40°) showing cracks. These deposits are for the droplets that are dried at room temperature but with RH (a) 20%, (b) 40%, (c) 60% and (d) 90%. The number of cracks can be seen to decrease with the increase in RH. The base diameter of each drop is around 3 mm [73].

***Mechanical property of the colloidal deposit*** – Since the critical strain energy release rate ($G_c$) depends on the material properties such as $E_f, \nu_f, K$ of deposit, the nucleation of cracks also depends on these quantities. In principle, the magnitude of these quantities viz. $E_f, \nu_f, K$ in a desiccating colloidal deposit is always modulated, but towards the tail end of drying i.e. when the crack nucleate they are nearly constant. Tirumukudulu et al. has derived expression of critical stress $\sigma_c$ for a thin colloidal deposit which is allowed to dry isotropically. It is expressed as [59,75],

$$\frac{\sigma_c R_p}{2\gamma} = 0.1877 \left(\frac{2R_p}{h}\right)^{2/3} \left(\frac{G(E_f, \nu_f) M \Phi_{rcp} R_p}{2\gamma}\right)^{1/3} \tag{19}$$

where, $R_p$ = radius of constituting particles,
$\gamma$ = surface tension of air-water interface,
$M$ = coordination number,
$G(E_f, \nu_f)$ = shear modulus,
$h$ = thickness of the colloidal deposit.

Please note that the critical stress expressed in Eq. 19 was derived Routh-Russel model. The corresponding strain energy release rate is obtained from an expression of critical stress. In experiments, the mechanical property of colloidal deposit are manipulated either by changing the constituents or by varying the random close packing fraction ($\Phi_{rcp}$) of the particles or by varying the drying rate ($V_E$).

†‡Contributed equally
*Corresponding author:hisaylama@gmail.com

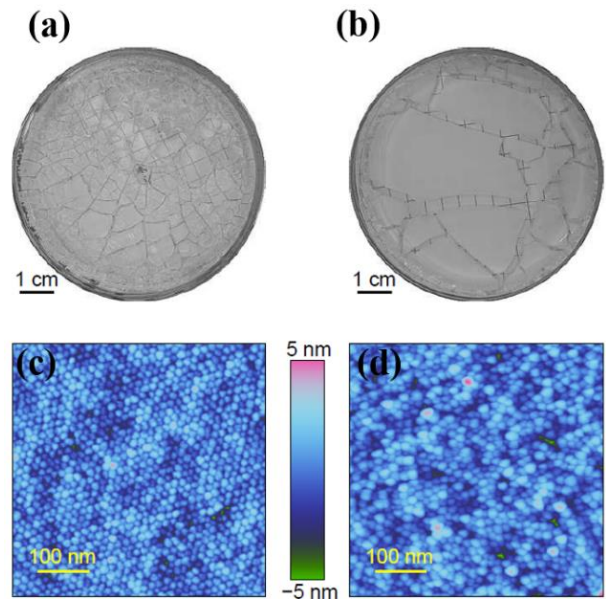

Figure 21: The picture of dried deposits on a petri dish with cracks. The deposit comprises of Ludox® HS-40 with an initial volume fraction of $\sim 0.4$. The picture of dried deposit with cracks obtained by drying (a) at RH $\sim 10\%$ and (b) RH $\sim 95\%$. (c), (d) Atomic force microscopy (AFM) images for the respective deposits depicts their local micro-structure [74].

***Microstructure*** – The particle micro-structure is an important parameter that dictates the crack pattern in the colloidal deposit. For example, Priorid et al. [74] has reported the correlation between the number density of cracks (number of crack per unit length) and the local microstructure i.e. the arrangement of the particles. The crack-density was high when the particles in a deposit are hexagonally close-packed, and it was significantly low when the constituting particles were randomly arranged (shown in Fig. 21(a)-(b)). In their experiments, the particle arrangement was modulated by varying the drying rate $V_E$. High drying rate was found to result in the ordered closely packed assembly of particles (shown in Fig. 21(c)) while the slower drying rate yields an amorphous phase (shown in Fig. 21(d)). Please note that these variations were obtained for the deposits comprising of spherical particles. For a deposit with non-spherical particles, the local arrangement of particles can vary between an isotropic phase and the nematic phase [76]. The degree of order or disorderedness in the particle arrangements can be characterized by a quantity termed 2D orientation order parameter $S$ and is expressed as [80],

$$S = <cos2\psi> \tag{20}$$

†‡Contributed equally

*Corresponding author:hisaylama@gmail.com

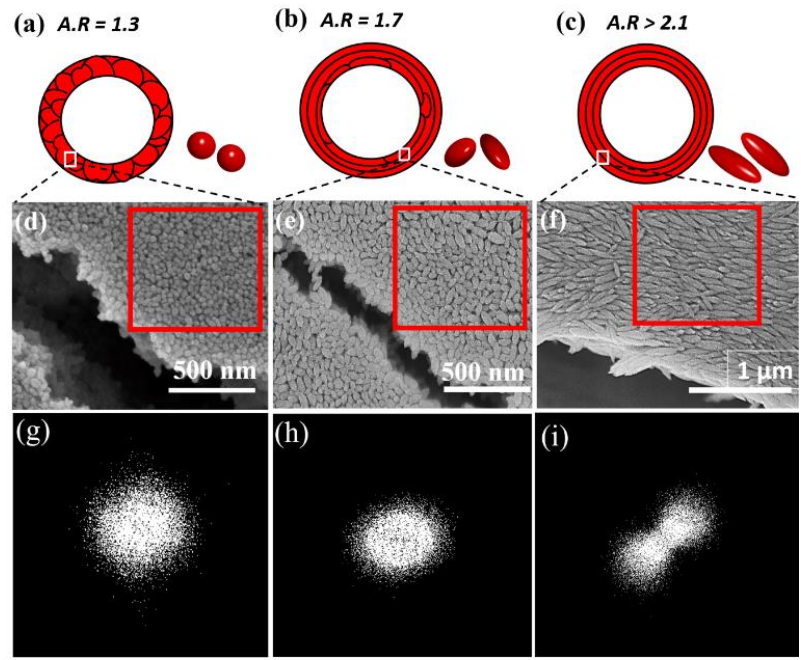

Figure 22: (a) - (c) Schematic of the dried ring like deposit obtained by drying colloids (hematite ellipsoids, pH ≈ 2) in sessile drop geometry at T$_{sub}$ ≈ 25°C and RH ≈ 40%. Scanning electron microscopy (SEM) images showing the region in the vicinity of cracks for the deposit comprising of particles with SEM micrographs depict the particle assembly in the vicinity of the cracks for the deposits with particle aspect ratio with (d) AR ≈ 1.3, (e) AR ≈ 1.7, and (f) AR > 1.3. (g) − (i): The fast Fourier transform (FFT) pattern of the deposit surface in the region marked by the rectangle in the respective SEM micrographs. The clear transition in the micro-structure from an isotropic phase to anisotropic phase resulting in the variation of crack morphology from radial or circular is observed [76].

where $\psi$ is an angle between major axis of a nonspherical particle and the reference axis. The value of $S$ varies between $0 - 1$, for perfectly isotropic assembly of particles $S = 0$, for nematic like phase $0.4 < S < 0.9$ while $S = 1$ for perfect crystals. For the drying deposit with non-spherical particle, the degree of orderedness was controlled by varying the aspect ratio (AR) of the constituting particles. Lama et al. [76] have reported that the deposit with particles of AR > 2.1 exhibits nematic like ordering while that with AR < 1.7, particles are randomly arranged. The resultant crack morphology was found to be significantly different (shown in Fig. 22(a)-(i)). The deposit with particles of AR > 2.1 was found to exhibit circular crack, while the deposit with AR < 1.7 radial or interconnected crack was observed.

***Cracks under external field*** – Desiccation cracks under an external field are one of the modular routes to generate the long-range spatial periodicity in cracks. The periodicity in cracks is obtained via the field-driven rearrangement of particles in the deposit. The external field typically enforces the particles to alter their orientation along the field. The selection of an external field to drive these particles essentially requires them to respond to the field. In the laboratory, magnetic field [81–83] and the electric field [84–86] are very common options that are exploited to the drive the particles in the dispersion. The typical experimental set-up for adopted for desiccation under magnetic and electric are depicted in Fig. 23(a)-(b).

†‡Contributed equally
*Corresponding author:hisaylama@gmail.com

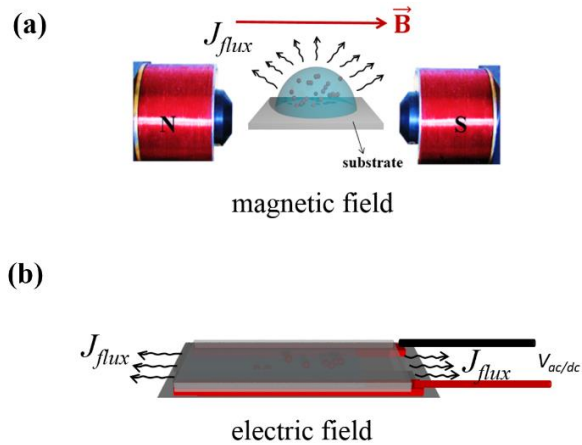

Figure 23: Schematic of standard drying geometry with an external field - depicts the drying of colloidal dispersion (a) under an in-plane dc magnetic field $\vec{B}$ and (b) under an electric field $\vec{E}$. (Images borrowed from PhD. thesis of Hisay Lama submitted to IIT Madras).

The colloid comprising of magnetic particles under the uni-axial dc magnetic field $\vec{B}$ tend to reorient them such that they form a chain-like structure. Customarily, the cracks in a deposit comprising of spherical magnetic particles are always observed to align along the direction of an applied field. For a deposit comprising of magnetite ($\gamma$−Fe$_2$O$_3$) or hematite ($\alpha$−Fe$_2$O$_3$) particles, under a uniaxial dc field is reported to exhibit parallel cracks that are parallel to the direction of $\vec{B}$ [83]. The magnetic particles in the dispersion, typically possess the intrinsic magnetic moment $\vec{\mu}$ and interact with each other. Upon an application of dc magnetic field $\vec{B}$, the particles with an intrinsic magnetic moment $\vec{\mu}$ reorient themselves such that $\vec{\mu} || {}^l\vec{B}$ as shown in Fig. 24(a). The net magnetic field-driven energy experienced by the particles is given by,

$$U_M = \Sigma_s \overrightarrow{\mu_s} . \vec{B} + U_{int}^M \tag{21}$$

where '$s$' is the number of particles and $U_{int}^M$ is the dipolar interaction energy. When $\vec{B}$ is sufficiently high, such that $U_M$ is greater than the thermal energy and hydrodynamic driven effects, particles exhibit magnetic field-driven alignment followed by the change in orientation of cracks.

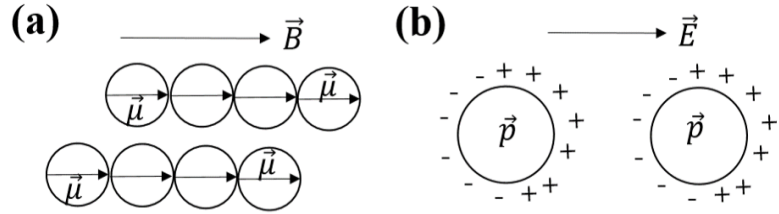

Figure 24: Schematic shows the external field-driven alignment of spherical particles. (a) Alignment of magnetic particles with an intrinsic magnetic moment $\vec{\mu}$ under a magnetic field $\vec{B}$. (b) Schematic of charged particles under an electric field $\vec{E}$ induces dipole moment $\vec{p}$. (Images are borrowed from PhD. thesis of Hisay Lama submitted to IIT Madras).

Similarly, the desiccation under the electric field is another popular technique that is exploited to generate spatially regular cracks [87,88]. The cracks are reported to align along the direction of electric lines of force. The application of electric field in the dispersion comprising of charged particles, typically manipulate their arrangement. These particles interact with each other via electrostatic interactions. When an electric field, $\vec{E}$ is applied to the dispersion, constituting charged particles gets electrically polarized and the charge distribution on the surface becomes asymmetrical with the particles possessing dipole moment $\vec{p}$. The net energy associated with the dispersion is expressed as,

$$U_E = \Sigma_s \vec{p_s} \cdot \vec{E} + U_{int}^E \tag{22}$$

where, '$s$' is the number of particles and $U_{int}^E$ the interaction energy between the dipoles. Analogous to the magnetic field, when $U_E$ is greater than thermal and hydrodynamic driven effects, particles and the resultant cracks both undergo field-driven alignment.

**SUMMARY**

To summarize, we presented a brief overview of the physics of drying with the primary focus on describing the phenomenon of consolidation, the formation of dried pattern and nucleation of desiccation cracks. The topics discussed in this chapter are pedagogical and are relevant to understand various physical phenomena such as self-assembly, pattern formation and desiccation cracks.

†‡Contributed equally
*Corresponding author:hisaylama@gmail.com

Drying of colloid and the formation of particle deposit can be seen as a process in which their physical state transforms from a liquid to a solid deposit. This happens mainly via evaporation induced diffusion of vapour into the ambient atmosphere and the accumulation of particles due to the local flow field. We explain these flow fields for a model drying geometry namely sessile drop. Thereafter, the formation of dried patterns such as coffee-ring, uniform deposit and multiple rings are illustrated and their correlation on various physical parameters are also explained. Also, we have shown some recent results that underline the geometrical confinement induced patterning of colloids.

The initial process of drying and their consolidation into a solid deposit results in the generation of tensile stress and such deposit undergoes deformation leading to an accumulation of strain energy. The phenomenon of development of such strain energy and its release for nucleating cracks are discussed briefly in this chapter. The nucleation of cracks and their organization into various morphology is found to depend on numerous parameters, such as physicochemical conditions of drying, the mechanical property of particle deposit and the rearrangement of local microstructure. Besides, external field-driven rearrangement of particles in the deposit and their impact on cracks is also discussed.

Tailoring the dried particle deposit and the resultant cracks into a suitable pattern have potential applications in material development e.g. photonic crystals and lithographic templates. Therefore, to gain control of these specific applicability's and their manipulation via drying driven phenomena requires a thorough understanding of the various theoretical and experimental aspects relating to these subject. In that way, we believe that this chapter serves the purpose and provides an introductory background on the topic.

**ACKNOWLEDGEMENT**

We thank Prof. Dillip K. Satapathy (Dept. of Physics IITM) and Prof. Madivala G. Basavaraj (Dept. of Chemical Engineering) for useful suggestions in writing and editing this research notes.

†‡Contributed equally

*Corresponding author:hisaylama@gmail.com

†‡Contributed equally
*Corresponding author:hisaylama@gmail.com

†‡Contributed equally
*Corresponding author:hisaylama@gmail.com

†‡Contributed equally
*Corresponding author:hisaylama@gmail.com

[†‡]Contributed equally

*Corresponding author:hisaylama@gmail.com